# EVIDENCE AGAINST KLEIN PARADOX IN GRAPHENE

D. Dragoman – Univ. Bucharest, Physics Dept., P.O. Box MG-11, 077125 Bucharest,

Romania, e-mail: danieladragoman@yahoo.com

**Abstract**

It is demonstrated that both transmission and reflection coefficients associated to the Klein

paradox at a step barrier are positive and less than unity, so that the particle-antiparticle pair

creation mechanism commonly linked to this phenomenon is not necessary. Because graphene

is a solid-state testing ground for quantum electrodynamics phenomena involving massless

Dirac fermions we suggest that the transport characteristic through a *p-n* graphene junction can decide between the results obtained in this paper and the common Klein paradox theory, which imply negative transmission and higher-than-unity reflection coefficients. Recent experimental evidence supports our findings.

## 1. Introduction

The Klein paradox refers to the propagation of a relativistic quantum particle, described by the Dirac equation, through a sufficiently high potential barrier without exponential dumping. Klein himself showed that, although the potential barrier becomes transparent if high enough, the transmission and reflection coefficients are still less than unity; this behavior can be understood by a proper consideration of the role of negative-energy solutions of the Dirac equation in the charge transport process. However, later authors claimed that the reflection coefficient at the step barrier is higher than 1 and the transmission coefficient is negative, phenomenon associated with creation of electron-positron pairs at the potential discontinuity. Since most of the controversies related to the Klein paradox are connected to the explanation of the seemingly unphysical negative transmission and higher-than-unity reflection coefficients, these features can be regarded as the real Klein paradox. In particular, throughout this paper, we refer to these features as the Klein paradox. A review of the historical development of the Klein paradox can be found in [1].

Astonishingly, when a finite-width barrier, instead of a step-like barrier, is considered, the reflection and transmission coefficients become well behaved (in the sense of being positive and less than 1), in apparent disagreement with the electron-positron pair creation at the step barrier. A probable pair annihilation phenomenon at the other potential discontinuity has been imagined to account for this result [2].

It is important to emphasize that no direct experimental evidence of the Klein paradox has been found yet. However, recent results demonstrate that charge carriers in graphene have a massless (and hence gapless) Dirac-like dispersion relation, fact that allows solid-state testing possibilities of quantum electrodynamics phenomena, including the Klein paradox [3]. In particular, unit transmission coefficient at normal incidence on a finite-width potential barrier in graphene has been already predicted [3]. It becomes feasible therefore to test the Klein paradox at a step-like potential discontinuity. If particle-antiparticle (in this case electron-hole) pair production at the interface were assumed, such an experiment would be expected to produce a net current that flows in an opposite direction to the electron current incident on the step barrier. But no associated current source can be identified (the corresponding source of electron-positron pair generation in quantum electrodynamics is the quantum vacuum). The possibility that graphene is not a solid-state analog of massless Dirac fermions has been ruled out by both theory [4-5] and experiments [6-8].

We are thus faced with a dilemma: either the Klein paradox does not occur or an unknown current mechanism exists in graphene that would produce apparent odd results. Disregarding the second possibility as unphysical, the aim of this paper is to show that, indeed, there is no Klein paradox. More precisely, we demonstrate theoretically that the reflection and transmission coefficients of a step barrier are both positive and less than unity, and that the hypothesis of particle-antiparticle pair production at the potential step is not necessary. Recent experimental evidence that supports our work is indicated [9].

## 2. The Klein paradox revisited

The plane-wave positive- and negative-energy solutions of the relativistic time-independent Dirac equations, $\Psi_+ = \exp[-iEt + i\boldsymbol{p} \cdot \boldsymbol{x}]\psi_+(\boldsymbol{p})$ and $\Psi_- = \exp[iEt - i\boldsymbol{p} \cdot \boldsymbol{x}]\psi_-(\boldsymbol{p})$, respectively, are four-dimensional column vectors that satisfy the equations [10]:

$$\left[i\left(\beta\frac{\partial}{\partial t} + \beta\boldsymbol{\alpha} \cdot \nabla\right) - m\right]\psi_+ = 0 , \quad \left[i\left(\beta\frac{\partial}{\partial t} + \beta\boldsymbol{\alpha} \cdot \nabla\right) + m\right]\psi_- = 0 . \tag{1}$$

Although the form of these solutions is well known, in many textbooks and papers errors often do occur, so that for clarity and completeness we give here their expressions:

$$\psi_{+\uparrow} = N_+ \begin{pmatrix} E+m \\ 0 \\ p_z \\ p_x + ip_y \end{pmatrix}, \quad \psi_{+\downarrow} = N_+ \begin{pmatrix} 0 \\ E+m \\ p_x - ip_y \\ -p_z \end{pmatrix}, \quad \psi_{-\uparrow} = N_- \begin{pmatrix} p_z \\ p_x + ip_y \\ E-m \\ 0 \end{pmatrix}, \quad \psi_{-\downarrow} = N_- \begin{pmatrix} p_x - ip_y \\ -p_z \\ 0 \\ E-m \end{pmatrix}.$$

(2)

The $\uparrow, \downarrow$ subscripts refer to states with spin-up and spin-down, respectively, the normalization factors are $N_{\pm} = \{2\pi[2E(E \pm m)]\}^{-1/2}$, $E$ is the electron energy, $\boldsymbol{p} = (p_x, p_y, p_z)$ its momentum, and the dispersion relation is $E^2 = \boldsymbol{p}^2 + m^2$, with $m$ the electron mass. The light velocity in vacuum is taken as $c = 1$. We consider in the following one-dimensional propagation, say in the $z$ direction, and absence of spin-flipping processes, case in which the solutions of the Dirac equation take the simpler, spinor form

$$\psi_{\pm}(p) = \begin{pmatrix} p \\ E-m \end{pmatrix},$$

(3)

which must be multiplied with the factor $N = \{2\pi[2E(E-m)]\}^{-1/2}$ to ensure proper normalization.

Consider now the situation depicted in Fig. 1, in which an electron is incident from the left on the step barrier. The total wavefunction in region I is a sum of the incident and reflected waves: $\Psi_I = N_I \exp(-iEt)[\psi_+(p)\exp(ipz) + r\psi_+(-p)\exp(-ipz)]$, with $p = (E^2 - m^2)^{1/2}$. In region II, in the Klein energy interval, for which $V_0 > E + m$, negative-energy electron states are involved in the tunneling process so that the wavefunction is $\Psi_{II} = N_{II} \exp[i(E-V_0)t]\psi_-(-q)\exp(iqz)$, with $q = [(E-V_0)^2 - m^2]^{1/2}$. The wavefunction normalization coefficients in regions I and II are, respectively, $N_I = \{2\pi[2p(E-m)]\}^{-1/2}$ and $N_{II} = \{2\pi[2q \mid E - V_0 - m \mid]\}^{-1/2}$. Note that in region II we have considered a negative-energy electron state that propagates in the same direction as the incident positive-energy state and have therefore taken $\psi_-(-q)$ instead of $\psi_-(q)$. Also note that the electron energy in region II is negative: $E - V_0 < 0$, as it should be.

To obtain the reflection and transmission coefficients, $R$ and $T$, we must impose the continuity condition at the boundary $z = 0$, from which it follows that

$$R = \left( \frac{1 - \kappa}{1 + \kappa} \right)^2, \quad T = \frac{4\kappa}{(1 + \kappa)^2} \tag{4}$$

with

$$\kappa = \frac{-q}{p} \frac{E - m}{E - V_0 - m} = \sqrt{\frac{(V_0 - E - m)(E - m)}{(V_0 - E + m)(E + m)}} \ . \tag{5}$$

It is easy to show that in the Klein energy interval $0 \leq \kappa \leq 1$, so that $0 \leq R, T \leq 1$ and $R + T = 1$. Except for the fact that under the barrier $q$ is real, these values of $R$ and $T$ are typical for wavefunction refraction at a boundary and/or for tunneling processes, and do not imply exotic phenomena such as particle-antiparticle pair creation. As long as the electron energy spectrum is considered split into a positive- and a negative-energy part, separated by a forbidden region, there is no paradox.

Actually, the inverse of the $\kappa$ coefficient in (5), i.e. $1/\kappa$, was the corresponding coefficient in the original paper of Klein, reproduced in equation (6) in [1]. It is easy to see that both reflection and transmission remain the same when $\kappa$ is replaced by $1/\kappa$, which means that we have recovered the original expressions of Klein without following the same line of judgment. In particular, our choice of $-q$ instead of $q$ follows from the form of the Dirac solutions and not from considerations regarding the group velocity. (Actually, the group velocity, defined as $v_g = d(E - V_0) / dq$ or $v_g = dE / dq$, is positive for our choice of propagation direction.)

Remarkably, the expression of $\kappa$ in the common version of the Klein paradox, denoted here by $\kappa'$, differs from $\kappa$ in (5) (or rather from $1 / \kappa$) by just a negative sign: $\kappa' = q(E + m) / [p(E + m - V_0)]$. Klein introduced this negative sign ad hoc from considerations of antiparticle propagation direction [1]. As such, this motivation was probably not convincing and later authors dropped it, with the expense of introducing particle-antiparticle pair creation to account for the resulting negative transmission and the higher-than-unity reflection. Moreover, several inconsistencies and mistakes in the derivation of $\kappa'$ appear in different works, so that it is difficult to get a clear idea of its proper derivation. For example, in [10], the wavefunction in region II is taken to be the positive-electron wavefunction, although negative-energy states are clearly involved. It is easy to check that (5) would be obtained if the correct wavefunction in region II would be used and if the correct four-dimensional column vectors (2) would be employed. In another cited paper, [1], although the correct spinor forms of the Dirac equations are used (see equations (9-10), without the imaginary $i$ in the upper row), the wavefunction in region II (equation (15) in [1]) is wrong again. Last, but not least, in [2] the antiparticle propagation direction is the same as here, but there is an extra minus sign in the upper row of the four-dimensional column vector under the barrier (equation (2) in this paper), which leads to the wrong $\kappa'$. To avoid confusions, the correct Dirac solutions are included in this paper as equations (2) or (3), and can be verified by the reader. The demonstration of (5) should then be free from ambiguities. This implies, however, that the literature associated to the Klein paradox, and in particular the mechanism of particle-antiparticle pair creation at the potential discontinuity is without object.

### 3. The particle-antiparticle pair production phenomenon revisited

Although the phenomenon of particle-antiparticle pair production was considered as an explanation of higher-than-unity reflection coefficient and negative transmission at the step barrier in the Klein paradox, a single-particle description of the electron transport based on the Dirac equation is not appropriate to conclude about its existence. A many-body, quantum field theory is required, but this theory is based on the results of the single-particle theory. We show in this section that particle-antiparticle pair production is not necessary in order to explain the electron transport through a step barrier when the wavefunctions are chosen such that $R$ and $T$ are given by the expressions (4)-(5) in Section 2.

To start with, we define the particle wavefunctions $u_{\pm z}$ incident from the left (along the positive $z$ direction, subscript $+z$) and departing to the right (along the negative $z$ direction, subscript $-z$) without reflected wave as [11, 12]

$$
\begin{aligned}
u_{\pm z}(E, z) = & \, N_{\mathrm{I}} \frac{2\sqrt{\kappa}}{\kappa+1} \binom{\pm p}{E-m} \exp(\pm ipz)\theta(-z) \\
& + N_{\mathrm{II}}\theta(z)\left[ \frac{\kappa-1}{\kappa+1}\binom{\pm q}{E-V_0-m}\exp(\mp iqz) + \binom{\mp q}{E-V_0-m}\exp(\pm iqz) \right],
\end{aligned}
\tag{6}
$$

with $\theta(z)$ the Heaviside step function. Similar expressions hold for antiparticle wavefunctions departing to the left and incident from the right, denoted, respectively, by $v_{\pm z}(E, z)$:

$$
\begin{aligned}
v_{\pm z}(E, z) = & \, N_{\mathrm{I}}\theta(-z)\left[ \frac{1-\kappa}{\kappa+1}\binom{\mp p}{E-m}\exp(\mp iqz) + \binom{\pm p}{E-m}\exp(\pm iqz) \right] \\
& + N_{\mathrm{II}} \frac{2\sqrt{\kappa}}{\kappa+1}\binom{\mp q}{E-V_0-m}\exp(\pm iqz)\theta(z).
\end{aligned}
\tag{7}
$$

It can be easily seen that the currents corresponding to the solutions (6)-(7) are

$$j_{u\pm} = u_{\pm z}^+(E,z)\alpha_z u_{\pm z}(E,z) = \pm(2\kappa/\pi)/(\kappa+1)^2, \qquad (8)$$

$$j_{v\pm} = v_{\pm z}^+(E,z)\alpha_z v_{\pm z}(E,z) = \pm(2\kappa/\pi)/(\kappa+1)^2.$$

Although in [11] all four states in (6)-(7) have been used to define a wavefunction for the system in terms of creation and annihilation operators for particles and antiparticles incident from different directions, this is not necessary, since these states are not independent [10, 12]. In fact, only two of them are sufficient and, unlike in [11], both positive- and negative-energy states of the Dirac equation are needed in the quantum field theory, even in the energy range specific for the Klein paradox. More precisely, the quantum field operators can be expressed as either [10,12]

$$\psi(z) = \int dE[a_+(E)u_{+z}(E,z)\exp(-iEt) + b_+^+(E)v_{-z}(E,z)\exp(iEt)] \qquad (9a)$$

or

$$\psi(z) = \int dE[a_-(E)u_{-z}(E,z)\exp(-iEt) + b_+^+(E)v_{+z}(E,z)\exp(iEt)]. \qquad (9b)$$

In both cases, taking into account the anticommutation relations between the particle and antiparticle creation and annihilation operators: $\{a_\pm(E), a_\pm^+(E')\} = \delta(E-E')$ and $\{b_\pm(E), b_\pm^+(E')\} = \delta(E-E')$, as well as the expressions for the currents (8) it can be shown that the expectation value of the current associated to the field operator $\psi$ in the vacuum state is $\langle 0 | j_\psi | 0 \rangle = 0$. This result implies that no particle-antiparticle pair production must be associated with tunneling through a barrier step. The vanishing of the vacuum expectation value of the current in the presence of the potential step can be understood in terms of the vacuum states in [11] (see equations (26)-(29) in this reference) since the particle wavefunction incident from the right on the barrier and denoted by $u_R$ in [11] is in fact our $v_{-z}$ (this identification is also in agreement with [10-11]). Then, following the same line of judgment as in [11] and the same definition of the vacuum state, the vacuum expectation value of the current is found to vanish.

## 4. Klein paradox in graphene

Since the controversy related to the existence or not of the pair creation linked to the Klein paradox relies on minor and subtle differences in the derivation of the expressions of $R$ and $T$, it would be desirable to design experiments that would decide between the two possibilities: reflection and transmission given by (4)-(5), which does not rely on particle-antiparticle pair creation, and reflection and transmission with different expressions (in particular, with $\kappa'$ instead of $\kappa$ in (4)), which need the particle-antiparticle pair creation assumption to explain the results. Such an experiment is proposed in the following.

As stated above, graphene is a solid-state equivalent of a system of massless Dirac fermions. The results derived in Section 2 are valid for graphene if $m \to 0$. In particular, this means that the positive- and negative-energy states (corresponding in this case to electrons and holes) are no longer separated by an energy gap and therefore charge transport through a potential step occurs with no exponential dumping irrespective of the energy value.

This behavior is supported by the calculations in Section 2, which show that for $m \to 0$, $\kappa \to 1$, $R \to 0$ and $T \to 1$, i.e. the incident quantum wavefunction is totally transmitted across the step potential discontinuity. In contradistinction, if the Klein paradox theory in the common form were used, we would have $\kappa' \to -1$ in the $m \to 0$ limit, which means that $R$ and $T$ become infinite. More precisely, their value cannot be determined since their expressions become singular. Graphene is certainly a test case for the two theories.

Experimental results in favor of one or the other theory can be obtained with the configuration described in Fig. 2(a). The gate over half of the graphene sheet induces a potential barrier with a height dependent on the gate voltage $V_G$ applied through an isolating layer, while the two contacts are used to inject and collect charge carriers. To inject electrons normal to the step potential discontinuity, a bias voltage $V$ is applied between the two contacts. The carrier concentration in the graphene layer can be controlled through the voltage $V_b$ applied on a back gate, situated under the substrate layer in Fig. 2(a). If the theory presented in this paper is correct, the current through the exterior circuit has a direction consistent with the polarity of the applied voltage and intensity consistent with a transmission coefficient equal to unity. In fact, there is no difference in the *I-V* characteristic of this device between a graphene sheet and the device in Fig. 2(a) for normal electron incidence; differences appear only for oblique incidence. In particular, for a graphene sheet with carrier mobility $\mu = 15000 \text{ cm}^2\text{V}^{-1}\text{s}^{-1}$ and a carrier concentration $n = \alpha V_b$ with $\alpha = 7.3 \ 10^{10} \text{ cm}^{-2}\text{V}^{-1}$, the low-voltage and low-temperature *I-V* characteristics should be similar to those in Fig. 2(b), where the dotted-line, solid-line, and dashed-line curves correspond to $V_b = 0.1$ V, 0.2 V and 0.3 V, respectively. These graphene parameters are taken from [6]; a negative (positive) $V_b$ value induces hole-like (electron-like) carrier states. Recent experiments on a *p-n* junction in graphene confirm in fact the results of this paper [9]. The positive-energy carrier states, i.e. electrons in graphene, are transformed into negative-energy states, i.e. holes, when a potential barrier is applied, so that the *p-n* junction in graphene corresponds to a potential step in a Klein tunneling problem. The *p* and *n* regions in [9] are implemented and separately controlled by top and back gates on a graphene sheet, transport experiment results, in particular *I-V* curves in the absence of a magnetic field, being consistent with the simulations in Fig. 2(b).

It is interesting to note that these experimental results are in disagreement with calculations based on the graphene wavefunction obtained with the common theory at the basis of the Klein paradox. More precisely, for the wavefunction in [3] with spinor components (see Fig. 2(c) for notations)

$$\psi_1(x, y) = \begin{cases} \exp(ik_y y)[\exp(ik_{xI} x) + r \exp(-ik_{xI} x)], & x < 0 \\ \exp(ik_y y)t \exp(ik_{xII} x), & x \geq 0 \end{cases} \quad (10a)$$

$$\psi_2(x, y) = \begin{cases} \exp(ik_y y)s_I[\exp(ik_{xI} x + i\theta_I) + r \exp(-ik_{xI} x - i\theta_I)], & x < 0 \\ \exp(ik_y y)s_{II}t \exp(ik_{xII} x + i\theta_{II}), & x \geq 0 \end{cases} \quad (10b)$$

the amplitude transmission coefficient is $t = 2s_I \cos\theta_I / [s_I \exp(-i\theta_I) + s_{II} \exp(i\theta_{II})]$. Here $k_{xI} = k_F \cos\theta_I$, $k_y = k_F \sin\theta_I$ are the wavevector components in graphene, $k_{xII} = [(E - V_0)^2 / \hbar^2 v_F^2 - k_y^2]^{1/2}$, $\theta_{II} = \tan^{-1}(k_y / k_{xII})$, $k_F$ is the Fermi wavevector in graphene and $v_F \cong c / 300$ the Fermi velocity, $s_I = \text{sgn } E$, and $s_{II} = \text{sgn}(E - V_0)$. The dispersion relation in graphene is $E = \pm |\hbar \boldsymbol{k}| v_F$, where $\boldsymbol{k}$ is the electron wavevector. It is easy to see that for normal incidence, when $\theta_I = \theta_{II} = 0$, $t$ becomes infinite in the energy region $0 < E < V_0$. Although the transmission coefficient is singular and hence the *I-V* characteristics cannot be estimated in this case, the measured low-voltage conductance would have a modulus much higher than for the single-layer graphene, while the current through the exterior circuit would have an opposite direction to that imposed by the voltage source. This scenario is refuted by experiments [9], and is impossible since it would imply the existence of a hypothetical current source corresponding to the electron-hole pair creation, which, however, would generate energy that cannot be associated to any known physical mechanism; it would be as if a perpetuum mobile can be created.

On the contrary, according to our theory the spinor components of the wavefunction in the two regions should be

$$\psi_1(x,y) = \begin{cases} \exp(ik_y y)[\exp(ik_{xI}x) + r\exp(-ik_{xI}x)], & x < 0 \\ \exp(ik_y y)t\exp(-ik_{xII}x), & x \geq 0 \end{cases} \tag{11a}$$

$$\psi_2(x,y) = \begin{cases} \exp(ik_y y)[\exp(ik_{xI}x + i\theta_I) + r\exp(-ik_{xI}x - i\theta_I)], & x < 0 \\ \exp(ik_y y)t\exp(-ik_{xII}x - i\theta_{II}), & x \geq 0 \end{cases} \tag{11b}$$

case in which $t = 2\cos\theta_I /[\exp(-i\theta_I) + \exp(-i\theta_{II})] \to 1$ for normal incidence. In general, the transmission probability across the interface is $T = |t|^2 \cos\theta_{II} / \cos\theta_I$. It is interesting to note that (10) and (11) lead to the same expression of the amplitude transmission coefficient for tunneling across a potential barrier with finite width, instead of a potential step. This interesting result can be easily checked and shows that a pair annihilation mechanism is not needed to explain the transmission coefficient through a finite-width barrier; the negative sign in the exponent of the wavefunction in region II in our case balances the effects of the *s* factors and renders a correct transmission coefficient through a finite-width barrier while providing a plausible transmission coefficient through a potential step.

As mentioned previously, the potential discontinuity does not influence the current through the device in Fig. 2(a) for normal electron incidence, but its presence can be evidenced at oblique incidence. For example, for a barrier height of $V_0 = 0.3$ eV, and a Fermi wavelength of 50 nm, the angular dependence of the current, normalized to the value $I_0$ corresponding to normal incidence, is displayed in Fig. 2(d). The simulations have been made using the wavefunction in (11) and indicate a method to detect the presence of a potential discontinuity in graphene; a similar detection method for a finite-width barrier in graphene has been indicated in [3].

## 5. Conclusions

We have demonstrated that at particle tunneling in the Klein energy interval both transmission and reflection coefficients are positive and less than unity, so that the particle-antiparticle pair creation associated to the Klein paradox need not exist. The non-exponential damping of the incident quantum wavefunction is fully explained by the energy spectrum of the electron, which consists of positive- and negative-energy parts separated by an energy gap. Graphene, which is a solid-state testing material for quantum electrodynamics phenomena, can provide a definite answer about the true values of the transmission and reflection coefficients for the Klein paradox at a step barrier in the limit of massless Dirac fermions. An indirect result of the testing would indicate whether the particle-antiparticle pair creation phenomenon at a potential step must or must not occur. In this respect, an experimental configuration has been proposed, and recent experimental results that validate the theory presented in this paper have been indicated. We do not share the extended view in literature that the step potential is a pathological case, which leads to misleading conclusions. The fact that potential barriers are so often encountered in solid-state physics, especially in semiconductor heterostructures, indicates that their properties cannot be regarded as unphysical. Graphene is the only material in which such heterostructures are described by Dirac-like and not by Schrödinger-like equations; therefore step potentials in graphene should solve (and, in our opinion, have already solved through the experiments in [9]) the so-called Klein paradox.

**Figure captions**

Fig. 1   Energy diagram for the step barrier case of the Klein paradox.

Fig. 2   (a) Experimental configuration that can decide upon the correct formalism of the Klein paradox and (b) the *I-V* characteristic consistent with our treatment of the Klein paradox. (c) Top view of the device in (a), and (d) the angular dependence of the current through a potential discontinuity.

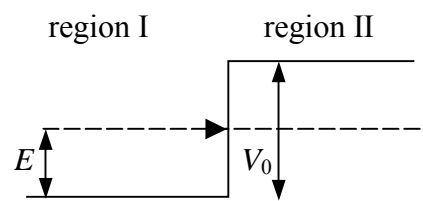

Fig. 1

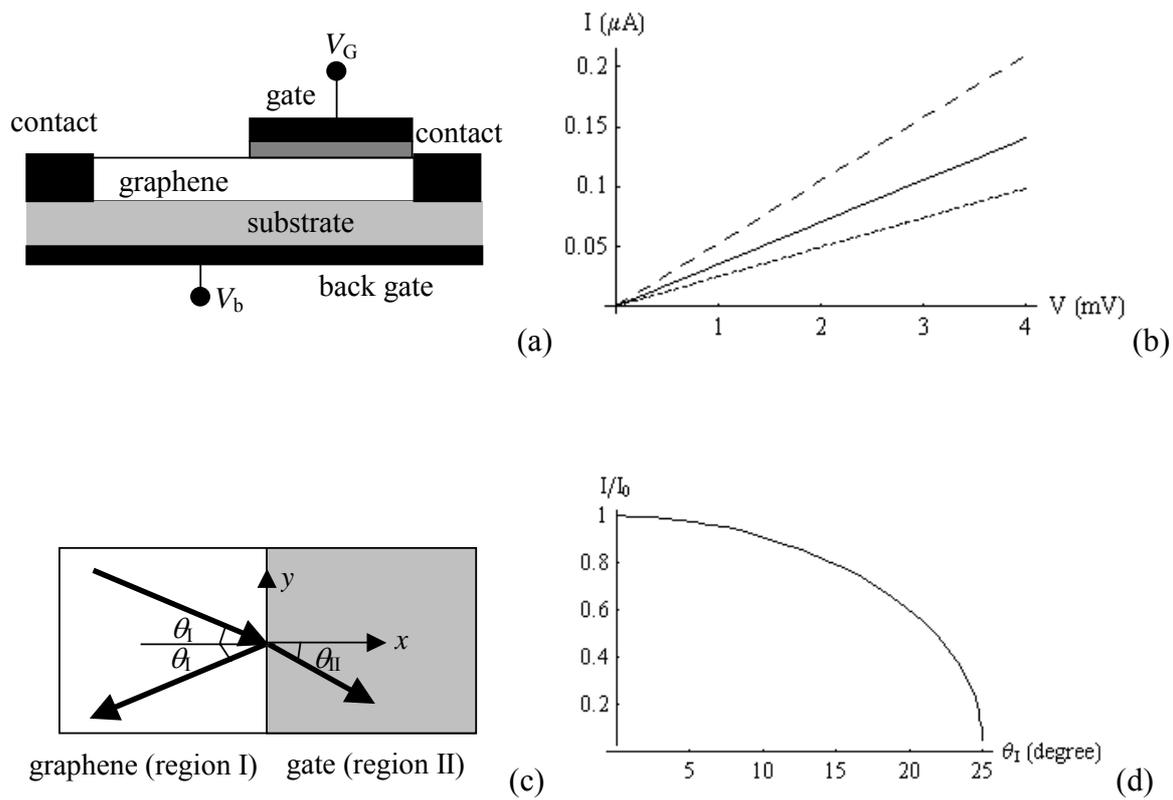

Fig. 2